# Evaluating noises of fast-simulated boson sampling with statistical benchmark methods

Yang Ji[1,2], Yongjin Ye[2], Qiao Wang[2], Shi Wang[1,2], Jie Hou[1,2], Yongzheng Wu[*1,2], Zijian Wang[2], Bo Jiang[*2]

[1]*Shanghai Research Center for Quantum Sciences, Shanghai 201315, China*

[2]*The 32nd Research Institute of China Electronics Technology Group Corporation, Shanghai 201808, China*

*Corresponding authors. E-mails: 782613169@qq.com, b26jiang@126.com

## Abstract

It is important to know noise levels of boson sampling in order to cautiously demonstrate the quantum computational advantage or realize certain tasks. Based on those statistical benchmark methods such as the correlators and clouds, which are initially proposed to discriminate boson sampling and other mockups, we quantificationally evaluate noises of photon partial distinguishability and photon loss compensated by dark counts. This is feasible owing to the fact that the output distribution unbalances are suppressed by noises, which are actually results of multi-photon interferences. This is why the evaluation performance is better when high order correlators or correspondent clouds are employed. Our results indicate that the statistical benchmark methods can also work in the task of evaluating noises of boson sampling. An effective scheme is also introduced to fast simulate noisy samples, especially those with photon partial distinguishability.

**Keywords**: boson sampling, simulation, statistical benchmark method, photon partial distinguishability, photon loss, dark count

## 1 Introduction

The proposal of quantum computing systems is challenged by realistic noises, preventing them from working correctly in a large scale. As an essential light-based candidate among these systems, boson sampling has attracted great attentions because it is free from constructing quantum logic gates [1, 2]. Until now, boson sampling has developed to demonstrate the so-called quantum computational advantages [3, 4], or realize certain tasks using its variants [5-10]. These are only guaranteed by the condition that realistic noises are extremely low, otherwise output deviations will erase multi-photon interference traces and boson sampling will degenerate to be approximately classical [11-19]. In these cases, quantum advantage claims are unreliable even though the output photon number is large.

The main noises in boson sampling are photon partial distinguishability [11, 16, 20] and photon loss [12, 16, 17, 21]. The former is due to intrinsic photon differences such as those of locations and frequencies during preparations of single-photon sources, while the latter is due to imperfections of practical devices such as linear interferometers. Photon loss is easy to judge by counting the output photon number, whereas it will be hidden if it is compensated by dark counts occurring in those detectors [22]. As a contrast, photon partial distinguishability is not intuitive. Both photon loss and partial distinguishability will suppress multi-photon interferences directly, resulting in the lack of quantum contributions in overall probability distributions and making boson sampling easier to be classically simulated. For example, considering the classical simulation algorithm owing to Clifford and Clifford [23], of which the computational complexity is $O[n2^n + \text{ploy}(m, n)]$, where $n$ and $m$ are output photon and mode numbers, it is easy to deduce that if the left photon number is scaled as $O(\log_2 n)$, then lossy boson sampling can be simulated by a classical computer in polynomial time [24]. As for photon partial distinguishability, the classical approximation algorithm using the cutoff interference photon number, owing to Renema [11], can simulate boson sampling within a little error of total variance distances, where only a finite interference scale is in need. Therefore, it is important to know noise levels, if one needs to estimate whether the results are owing to quantum processes or not.

Noises can be evaluated by calculating total variance distances of all the output patterns, which is computational inefficient [21]. Hence, it alternatively turns to those validation approaches [25-31] which initially only provide a binary judgement when dealing with boson sampling and other mockups, such as the mean-field sampling [32]. With this in mind, in our previous work [33], we extended the pattern recognition validation approach to evaluate noises of boson sampling, where both of the two main kind noises are considered. It is found that the extension is feasible based on the fact that the unbalanced distributions are suppressed by photon partial distinguishability, while the light-source occurring photon loss will worsen the validation performances because of the introduced uncertainty of correspondent submatrices. The unbalance is essential based on the fact that it will respond to the main kind noises in a monotonous way. This motivates us to explore the extension of other validation schemes which may be easier to conduct, or may reveal the noise information in a much more determinate routine.

In this paper, we focus on the statistical benchmark methods such as the correlators and the correspondent clouds [31, 34], which are also computational efficient. It is believed that those higher order correlators are naturally more sensitive to multi-photon interferences [33], which can be suppressed by noises. Moreover, the statistical benchmark methods land on the framework where the random matrix theory works [31]. Therefore, one does not need to learn the interferometer information, even though it is at the cost of much larger amounts of samples.

This hints us that extracting the noise information from numerous output samples may be relatively easy in boson sampling.

Before noise evaluations, an effective scheme is introduced to fast simulate those noisy samples, based on those simulation algorithms such as the Clifford-Clifford algorithm, on the condition that photons are partially indistinguishable specially, considering that generally noises will reduce the simulation complexity.

# 2 Simulations of noisy boson sampling

In boson sampling, the probability of an output pattern $|T\rangle = |T_1, \dots, T_m\rangle$ is determined by [1]

$$\mathrm{Pr}_{\mathrm{ideal}}(T) = \frac{|\,\mathrm{Perm}(\mathbf{M}^{S,T})|^2}{\prod_{i=1}^{m} S_i! \prod_{i=1}^{m} T_i!}, \tag{1}$$

where $|S\rangle = |S_1, \dots, S_m\rangle$ is the input Fock state and $\mathbf{M}$ is the interferometer matrix. $\mathbf{M}^{S,T}$ is the submatrix generated by selecting rows and columns of $\mathbf{M}$ according to input and output states. $\mathrm{Perm}(\cdot)$ denotes the matrix permanent, of which the computational complexity is $O(n2^n)$ using the best-known algorithm [24], if the complex matrix is Haar random. Output samples can be obtained with classical simulation algorithms such as the rejection [35], Markov Chain Monte Carlo [35], and Clifford-Clifford [23].

Here, three kinds of noises are discussed, which are photon partial distinguishability, photon loss and dark counts. The latter two kind noises are discussed together because they can compensate with each other and make the output photon number equal to that of the input. Moreover, if one observes a pattern with fewer photons, one still can't exclude the case that some photons come from dark counts. Hence, it is hard to tell noise levels by counting the output photon numbers, no matter post selecting is used.

As for photon partial distinguishability, pairwise noise information is collected by a tensor, in which case the output probability is [11]

$$\mathrm{Pr}_{\mathrm{pd}}(T) = \frac{\sum_{\sigma}(\prod_{j=1}^{n} a_{\sigma_j j})\,\mathrm{Perm}(\mathbf{M}^{S,T} \odot \mathbf{M}_{1,\sigma}^{S,T*})}{\prod_{i=1}^{m} S_i! \prod_{i=1}^{m} T_i!}. \tag{2}$$

$\sigma = \{\sigma_1, \dots, \sigma_n\}$ runs over all the permutations of $\{1, \dots, n\}$ and $a_{ij} = x_{\mathrm{ind}} + (1 - x_{\mathrm{ind}})\delta_{ij}$ is the tensor element. $x_{\mathrm{ind}}$ is the signature of distinguishability. When $x_{\mathrm{ind}} = 1$ photons are indistinguishable and when $x_{\mathrm{ind}} = 0$ photons are totally distinguishable. $\mathbf{M}_{1,\sigma}^{S,T}$ is a rearrangement of $\mathbf{M}^{S,T}$, where columns are rearranged according to $\sigma$. $\odot$ is the elementwise product.

It is possible to employ those well-known classical algorithms to exactly simulate boson sampling with photon partial distinguishability. As an instance, here we employ the Clifford-Clifford algorithm, which is usually used for ideal boson sampling [23], by noticing that $1-x_{\text{ind}}$ actually is equivalent to the probability of a photon becoming totally distinguishable compared to others and entering into a virtual mode through a virtual beam splitter, which is similar to Gaussian boson sampling cases [18], as shown in Fig .1. For the $i$-th input photon $(i=1,..,n)$, the extended $m(n+1)$-dimensional matrix $\mathbf{B}_i$ of the virtual beam splitter between the $i$-th (actual) mode and the $i(m+1)$-th (virtual) mode, is a revision of the identity matrix, where the revised 4 elements become $b_{i,i}=\cos\omega$, $b_{i,i(m+1)}=-\sin\omega$, $b_{i(m+1),i}=\sin\omega$, $b_{i(m+1),i(m+1)}=\cos\omega$. $\cos\omega$ is the reflectivity of the beam splitter, which is related with $x_{\text{ind}}$ via

$$\cos^2\omega = x_{\text{ind}}. \tag{3}$$

For $n$ partially distinguishable photons, $n$ individual virtual $m$-mode interferometers characterized by $\mathbf{M}$ are constructed in Fig. 1, along with $n$ virtual beam splitters characterized by $\mathbf{B}_1$,.., $\mathbf{B}_n$ respectively. Therefore, the overall optical network matrix, including those actual and virtual parts, can be written as

$$\mathbf{U} = \mathbf{B}_1 \dots \mathbf{B}_n(\oplus_{i=1}^{n+1}\mathbf{M}). \tag{4}$$

Indeed, $\mathbf{U}$ has already contained noise information of photon partial distinguishability. Correspondingly, the input state is rewritten as $|S'\rangle = |S_1,\dots,S_m,0,\dots,0\rangle$, with zero states as the input for those virtual modes. Based on these, the Clifford-Clifford algorithm can now be employed, where the remaining work is to collect those photons measured in the actual and virtual detectors at the mode terminals, labelled by the same number in Fig.1, into one actual detector for one measurement time. The virtual optical parts will not greatly enhance the computational complexity of the algorithm because it is only polynomial to the mode number.

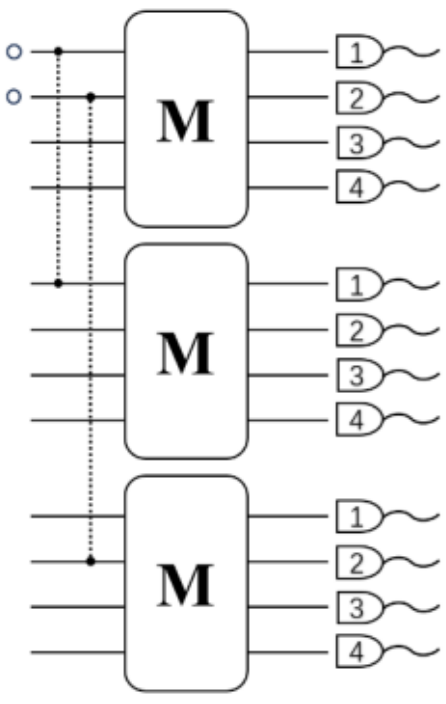

Fig. 1 Equivalent optical network of boson sampling with the noise of photon partial distinguishability. In this sketch, $n=2$ partially distinguishable photons enter into the interferometer characterized by $\mathbf{M}$ with $m=4$. The first interferometer is actual while the left two are virtual. The dot lines represent the virtual beam splitters between actual and virtual modes.

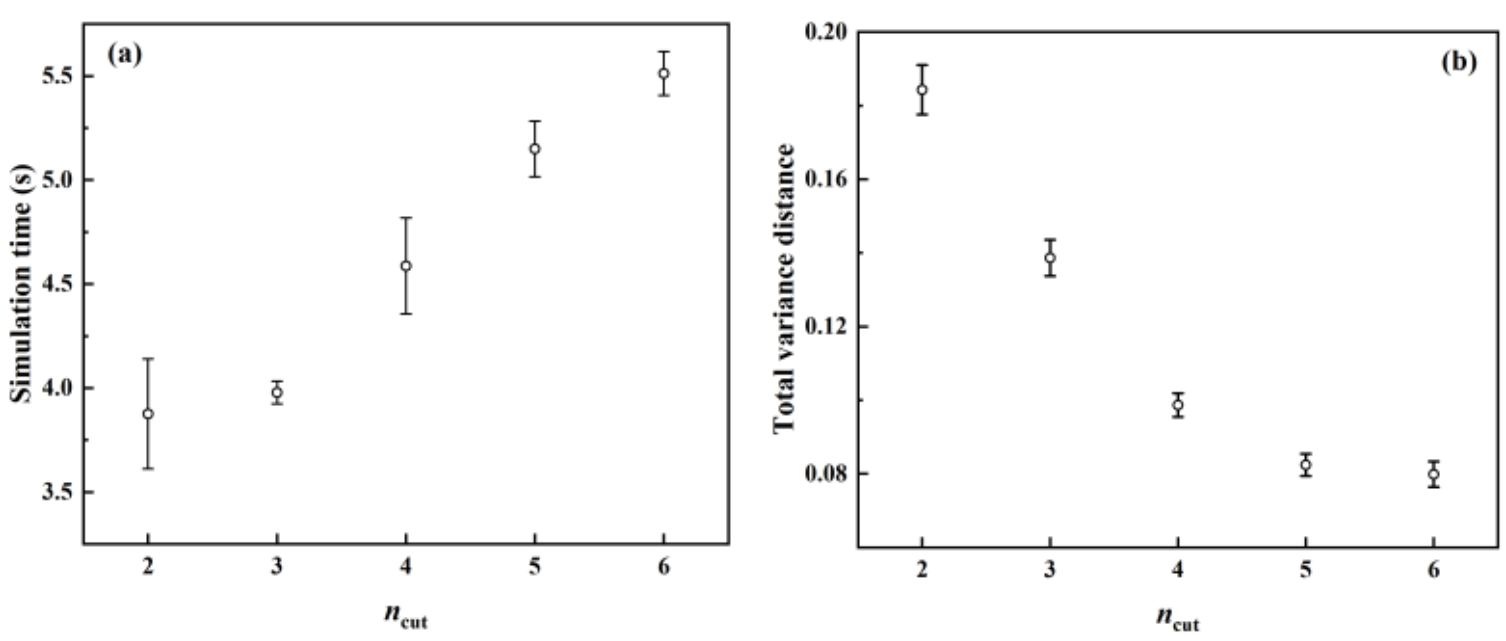

Fig. 2 Simulation results of samples with photon partial distinguishability, based on the approximate scheme, with changing the cutoff interfering photon number $n_{\mathrm{cut}}$. (a) Simulation time. (b) Total variance distances between the distribution of simulated samples and the theorical distribution. The parameters are $n = m = 6$, $x_{\mathrm{ind}} = 0.6$. In each test, for each $n_{\mathrm{cut}}$, the sample number is $10^4$. The repeat test time is 100. For each time, a random matrix is generated as the interferometer matrix. The bars present the standard deviations.

Equivalently and more importantly, simulations can also be carried out based on the Clifford-Clifford algorithm, using the scheme that for each input photon, it will be sent into the virtual interferometer characterized by $\mathbf{M}$ alone, with the probability of $1 - x_{\mathrm{ind}}$, where the simulation algorithm is then performed. Generally, the simulation is fast when $x_{\mathrm{ind}}$ is low because probabilities of high order multi-photon interferences are extremely low. Furthermore, the simulation can be faster if a cutoff interfering photon number is set, where those negligible high order multi-photon interferences will be ignored if $x_{\mathrm{ind}}$ is low. In detail, the correspondent approximate simulation procedure is presented in Algorithm 1. With this approximate scheme, the simulation time can be greatly reduced at the cost of total variance distances increased not significantly, as shown in Fig. 2.

---

**Algorithm 1** An approximate algorithm to generate a sample $\boldsymbol{T}$ for boson sampling with the noise of photon partial distinguishability, based on those simulation algorithms such as the Clifford-Clifford (CC) algorithm

---

**Require**: the input state written as $\boldsymbol{S} = (S_1, \dots, S_m)$, the interferometer matrix $\mathbf{M}$, the photon indistinguishability $x_{\mathrm{ind}}$ and the cutoff interfering photon number $n_{\mathrm{cut}}$

1: $\boldsymbol{S}_{\mathbf{ind}} \leftarrow \boldsymbol{S}$ // Initialize the input state with indistinguishable photons.

2: $\boldsymbol{L}_{\mathbf{dis}} \leftarrow \emptyset$ // Empty the list of all the input states with an individual photon.

3: **for** $i \leftarrow 1$ to $m$ **do**

4: **if** $S_i > 0$ **then**

5: **for** $j \leftarrow 1$ to $S_i$ **do**

6: **if** $\mathrm{Random}(0, 1) > x_{\mathrm{ind}}$ **then**

7: $\boldsymbol{S}_{\mathbf{ind}}[i] \leftarrow \boldsymbol{S}_{\mathbf{ind}}[i] - 1$

8: $\boldsymbol{S}_{\mathbf{dis}} \leftarrow \mathbf{0}_m$

9: $\boldsymbol{S}_{\mathbf{dis}}[i] \leftarrow 1$ // Send a distinguishable photon into a virtual mode.

---

10: $\boldsymbol{L}_{\mathbf{dis}} \leftarrow (\boldsymbol{L}_{\mathbf{dis}}, \boldsymbol{S}_{\mathbf{dis}})$ // List an input state with an individual photon.
11: **end for**
12: **end for**
13: **while** $\mathrm{Sum}(\boldsymbol{S}_{\mathbf{ind}}) > n_{\mathrm{cut}}$ **do** // Ensure at most $n_{\mathrm{cut}}$ interfering photons.
14: $\boldsymbol{S}_{\mathbf{ind}} \leftarrow \boldsymbol{S}$
15: $\boldsymbol{L}_{\mathbf{dis}} \leftarrow \emptyset$
16: **for** $i \leftarrow 1$ to $m$ **do**
17: **if** $S_i > 0$ **then**
18: **for** $j \leftarrow 1$ to $S_i$ **do**
19: **if** $\mathrm{Random}(0,1) > x_{\mathrm{ind}}$ **then**
20: $\boldsymbol{S}_{\mathbf{ind}}[i] \leftarrow \boldsymbol{S}_{\mathbf{ind}}[i] - 1$
21: $\boldsymbol{S}_{\mathbf{dis}} \leftarrow \mathbf{0}_m$
22: $\boldsymbol{S}_{\mathbf{dis}}[i] \leftarrow 1$
23: $\boldsymbol{L}_{\mathbf{dis}} \leftarrow (\boldsymbol{L}_{\mathbf{dis}}, \boldsymbol{S}_{\mathbf{dis}})$
24: **end for**
25: **end for**
26: **end while**
27: $\boldsymbol{T} \leftarrow \mathrm{Sample}(\boldsymbol{S}_{\mathbf{ind}}, \mathbf{M})$ // Generate a sample in actual modes with the CC algorithm.
28: **for** $i \leftarrow 1$ to $\mathrm{Len}(\boldsymbol{L}_{\mathbf{dis}})$ **do**
29: $\boldsymbol{T}_{\mathbf{dis}} \leftarrow \mathrm{Sample}(\boldsymbol{L}_{\mathbf{dis}}[i], \mathbf{M})$ // Generate a sample in virtual modes.
30: $\boldsymbol{T} \leftarrow \boldsymbol{T} + \boldsymbol{T}_{\mathbf{dis}}$ // Collect output photons in both actual and virtual modes.
31: **end for**
32: **return** $\boldsymbol{T}$

As for photon loss, a balanced loss model is considered, where for each mode, the photon transmittance is assumed to be constant [17]. This is suitable for a common symmetric interferometer with the optical unit number and the depth scaled as $O(m^2)$ and $O(m)$ respectively [36]. In this model, for each output photon, the loss probability is kept the same. The output photon number in each mode follows a binomial distribution and thus we can write down the output probability as

$$\Pr_{\mathrm{loss}}(T) = \sum_{T'}\{\Pr_{\mathrm{ideal}}(T')\prod_{i=1}^{m}[\binom{T_i'}{T_i}\eta_{\mathrm{t}}^{T_i}(1-\eta_{\mathrm{t}})^{T_i'-T_i}]\}, \quad (5)$$

where $T'$ runs over all the $n$-photon output patterns with $T_i' \geq T_i, \forall i$, and $n$ is now the input photon number. $\eta_{\mathrm{t}}$ is the transmission rate of each mode. Higher $\eta_{\mathrm{t}}$ corresponds to a lower photon loss probability.

Dark counts are only dependent on photon detector qualities and usually the correspondent dark count probability is quite low [22]. It may be necessary to consider this noisy effect when $m$ is extremely large. For example, in the collision-free regime where $m = O(n^2)$ [24], if the dark count probability $p_{\mathrm{dc}}$ is constant and the redundant photon number is scaled as $o(m)$, then it will be comparative to the effective photon number $n$ when $m$ is extremely large. Despite in general, dark counts are expected to have little influences on realizing the quantum computational advantage due to quite low $p_{\text{dc}}$, they will worsen those large-scale implement performances more or less. Note that with photon loss and dark counts, the photon number fluctuations will not enhance the computational complexity because the two kind noises have no influences on the background probabilities. Based on the assumption that the dark count probability is constant for each detector, and thus, the dark count detector number follows a binominal distribution, we can obtain the output probability via

$$\text{Pr}_{\text{dc}}(T) = \sum_{T'}[\text{Pr}_{\text{ideal}}(T')(1-p_{\text{dc}})^{m+n-\sum_i T_i} {p_{\text{dc}}}^{\sum_i T_i - n}], \tag{6}$$

where $T'$ now runs over all the $n$-photon output patterns with $0 \leq T_i - T_i' \leq 1, \forall i$. Writing down the output probabilities considering both photon loss and dark counts is complex because of numerous combinations. However, as for simulations, one only needs to change the original output patterns obtained based on Eq. (1), according to the binominal distributions mentioned above. Here the Clifford-Clifford algorithm is also employed to simulate those original samples. We are interested in those samples with $n$ photons because other samples can be excluded by post selecting.

As a supplement, the effective scheme to fast simulate samples with photon loss is also discuss briefly. Based on the assumption that photon loss occurs in light sources [21], the actual input combinations can be obtained, where input photons may be lost with a probability. Hence, the overall input photon number follows a binominal distribution. Since the computational complexity is determined by the actual input photon number, simulations can be greatly simplified if the photon loss probability is high. Actually, this scheme is only one part of that in the aforementioned photon partial distinguishability case, where those processes in virtual modes are ignored.

# 3 Noise evaluations

The statistical benchmark validation methods are extended to evaluate noises of those exactly simulated samples, by observing the mean photon number products of different mode combinations. More specifically, the $t$-th order correlator is expressed as [34]

$$\kappa(n_{o_1}, \dots, n_{o_t}) = \sum_\pi [(|\pi|-1)!\,(-1)^{|\pi|-1} \prod_{B\in\pi} \langle \prod_{i\in B} n_{o_i} \rangle]. \tag{7}$$

$\{o_1, \dots, o_t\} \subseteq \{1, \dots, m\}$ is the selected output mode number list without repetitions. $n_{o_i}$ is the photon number in the $o_i$-th mode. $\pi$ runs over all the partitions of $\{1, \dots, t\}$. According to Eq. (7), the 1st order correlator is $\langle n_i \rangle, i \in \{1, \dots, m\}$. This reveals little noise information because $\sum_i \langle n_i \rangle = n$ is fixed and there are only $m$ correlators, which are much fewer compared to those higher order correlators. More importantly, multi-photon interferences are more sensitive to noises [33], which are mainly revealed by higher order correlators. Fig. 3 shows the 2nd to 4th order correlators with changing the photon partial distinguishability. With increasing $x_{\text{ind}}$, those correlator points become closer to the line $x = y$. That is to say, the correspondent correlators become similar to the ideal ones gradually with increasing $x_{\text{ind}}$.

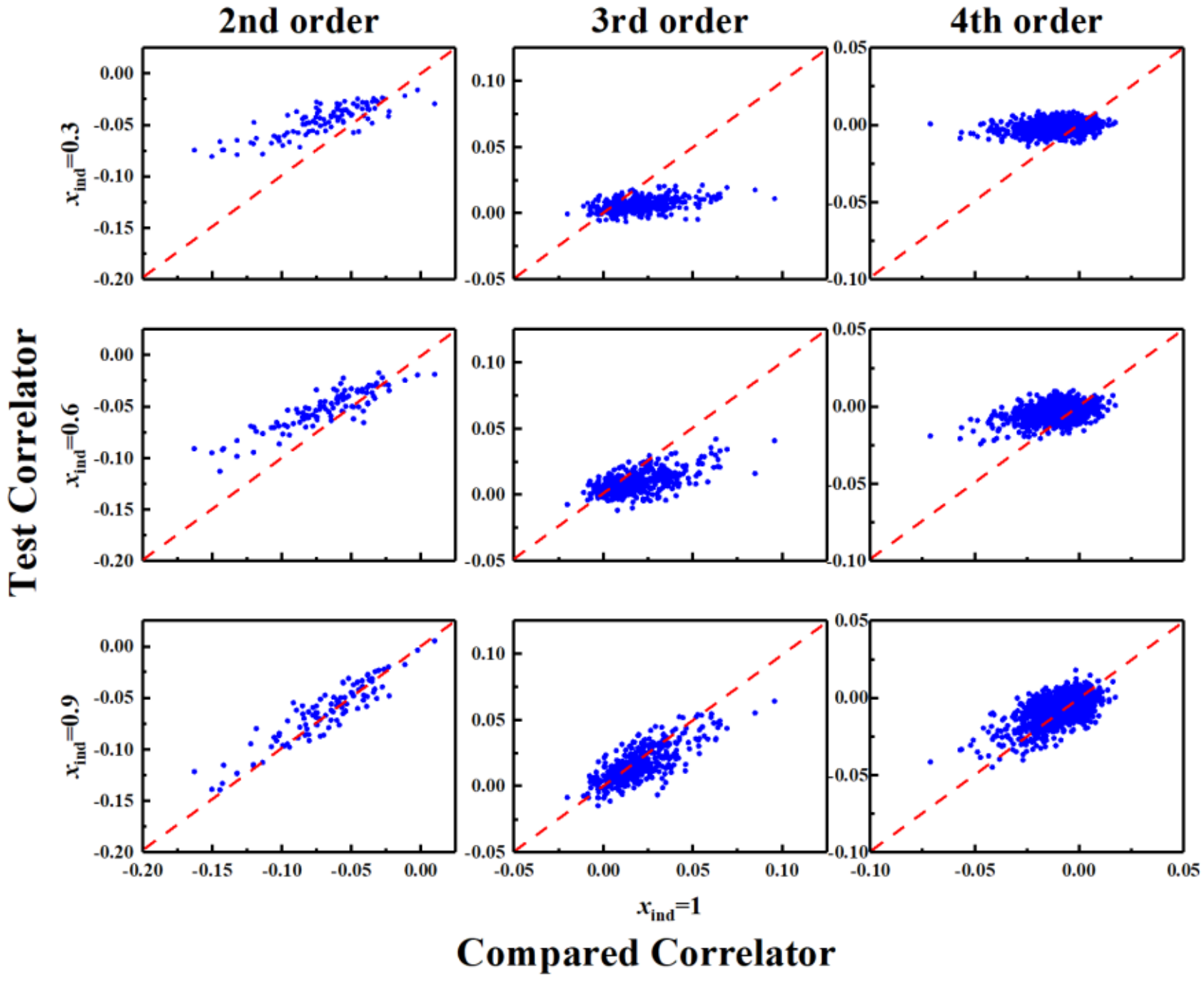

Fig. 3 Correlators for boson sampling with the noise of photon partial distinguishability. For each $x_{\text{ind}}$, $10^4$ samples are generated by simulation algorithms to obtain the correlators with $n = 10$ and $m = 15$. The red dash line is $x = y$.

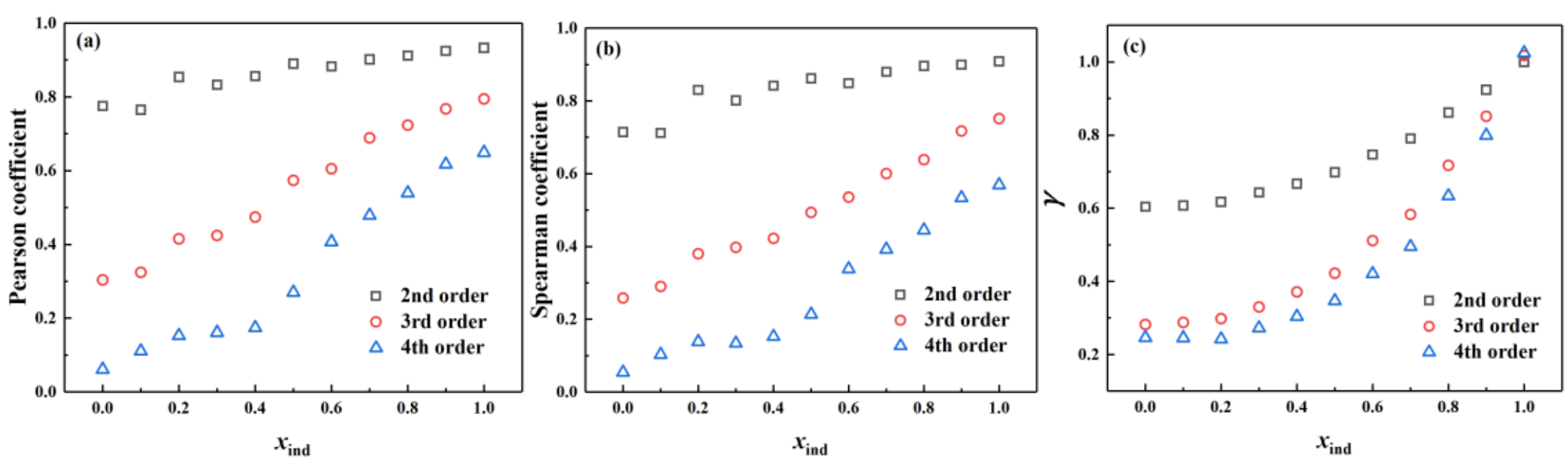

Fig. 4 Plots of relative coefficients and $x_{\text{ind}}$. (a) Pearson coefficients. (b) Spearman coefficients. (c) $\gamma$ reflecting point deviations.

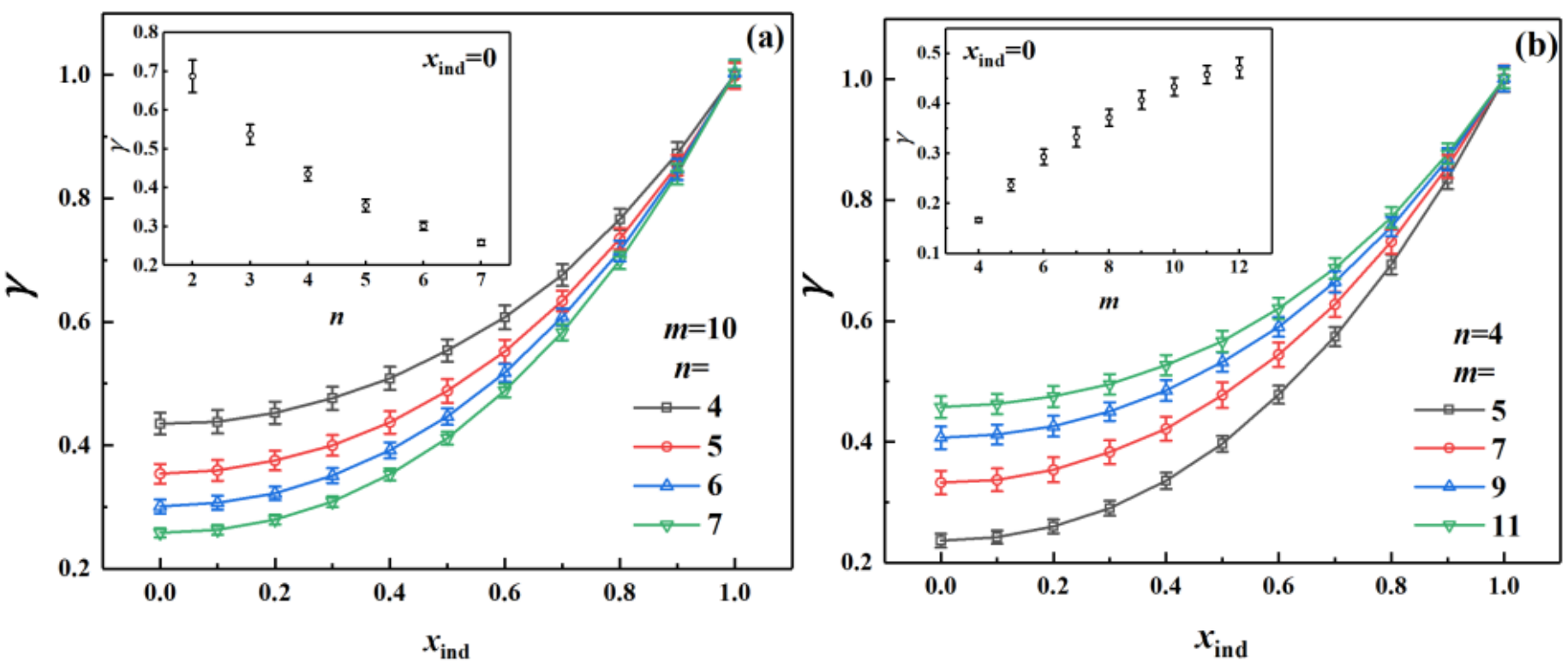

Fig. 5 Plots of $\gamma$ and $x_{\mathrm{ind}}$ with different scales, where the correlation order $t$ is 3. (a) $m$ is fixed and $n$ is changed. (b) $n$ is fixed and $m$ is changed. For each $x_{\mathrm{ind}}$, 100 random matrices are generated as the interferometer matrices. For each test, $10^4$ samples are generated. The insets show $\gamma$ with scales where $x_{\mathrm{ind}} = 0$. The bars present the standard deviations.

Noise information is firstly extracted by calculating Pearson and Spearman correlation coefficients. It is found that with increasing $x_{\mathrm{ind}}$, both the coefficients grow in tendency, which are more obvious with higher order correlators, as shown in Fig. 4. However, some points may drop slightly with $x_{\mathrm{ind}}$ even for those higher order correlators. This indicates that the correlation coefficients can only provide coarse evaluations of noises.

Alternatively, point deviations are investigated by calculating $\gamma = (\sum_i |\kappa_{\mathrm{test},i}|)/(\sum_i |\kappa_{\mathrm{comp},i}|)$, where $\kappa_{\mathrm{test},i}$ and $\kappa_{\mathrm{comp},i}$ are the $i$-th test and compared correlators. The correspondent tendencies are presented in Fig. 4 (c), where changes are obviously more continuous. Changes become sharp when photons are close to be indistinguishable. This is because only when $x_{\mathrm{ind}}$ is high, the multi-photon interferences become significant, otherwise the processes tend to be classical. Indeed, the higher order correlators provide better evaluation performances, whereas it is at the cost of computation time because for the $t$-th order, there are $\binom{m}{t}$ correlators in total to consider. Noise evaluations are affected by boson sampling scales. As shown in Fig. 5, with increasing $n$ or decreasing $m$, deviations become obvious for those samples with photon partial distinguishability when compared to the ideal ones. It is notable that with increasing $m$, $\gamma$ with $x_{\mathrm{ind}} = 0$ only grows asymptotically in the inset of Fig. 5 (b), indicating that the correlation method is robust even when $m$ is large.

The cloud method [31] is also extended by observing the coefficient of variation (CV) and the coefficient of skewness (CS), based on the correlators. This is believed to be free from learning the interferometer matrix information which usually is at the cost of extra experiments [37]. The results are alternatively made close to the expectation values according to the random matrix theory, through going over all the single-photon input combinations, in the regime $m \gg$ $n$. Fig. 6 presents the clouds formed based on the 2nd and 3rd order correlators. The clouds shift gradually with changing $x_{\text{ind}}$. This is more obvious for 3rd order correlators, indicating once again that the higher order correlators have a better evaluation performance. The mean CV and CS values of the clouds are shown in Fig. 6. The mean CV seems to be more sensitive to high $x_{\text{ind}}$ while the mean CS seems to be more sensitive to moderate $x_{\text{ind}}$.

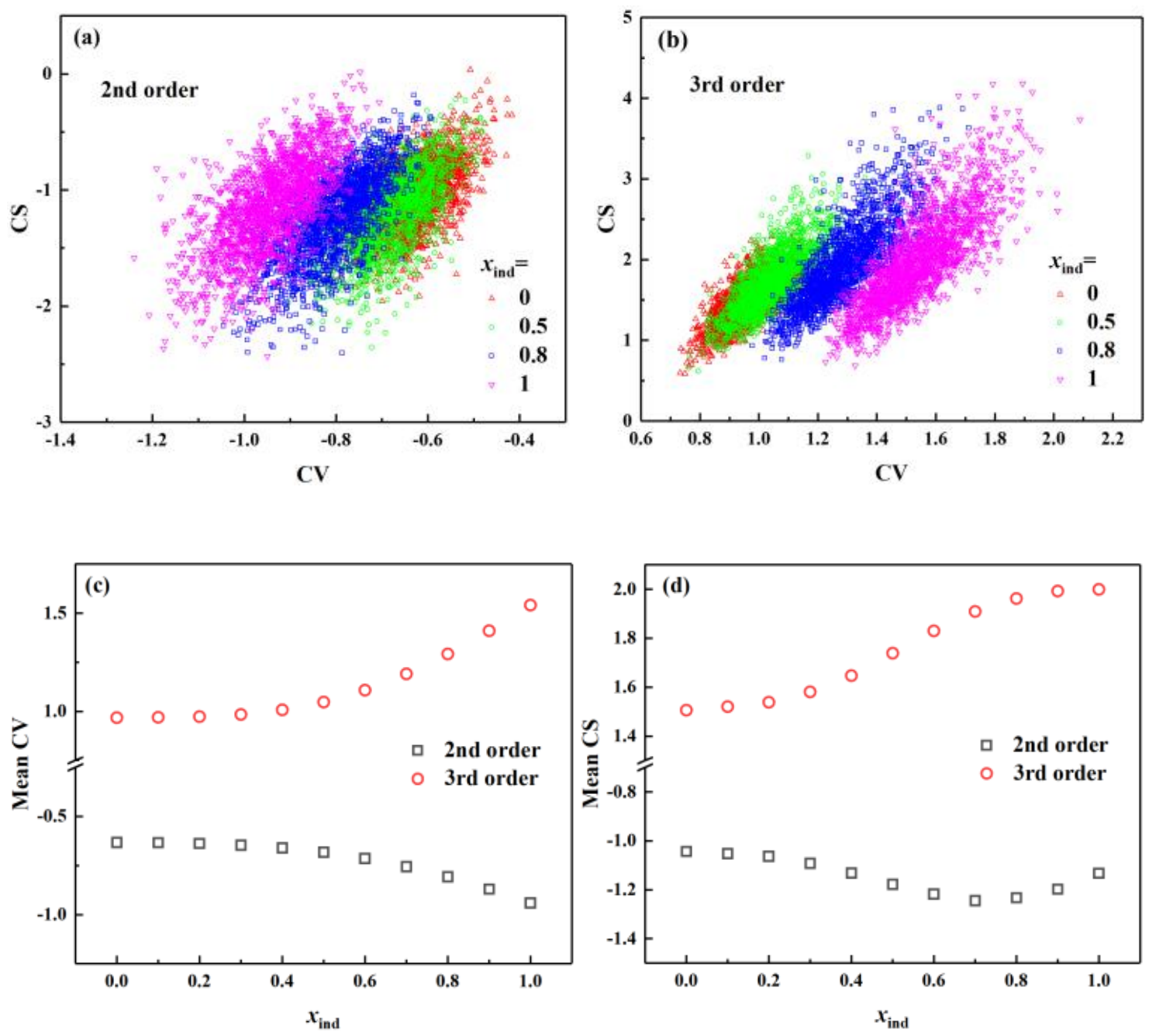

Fig. 6 The cloud method to evaluate the noise of photon partial distinguishability. (a)-(b) The clouds based on 2nd and 3rd order correlators, with $n = 4$ and $m = 16$. The sample number for each input combination and $x_{\text{ind}}$ is $10^4$. (c)-(d) Mean CV and CS values obtained based on the clouds with $x_{\text{ind}}$.

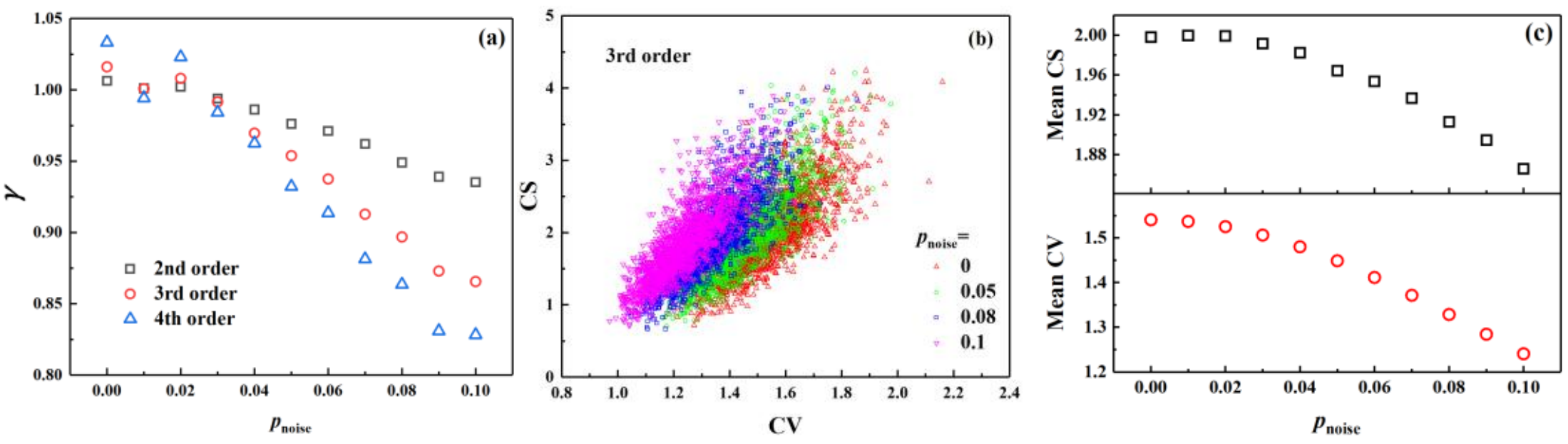

Fig. 7 The extended statistical benchmark methods for boson sampling with photon loss compensated by dark counts. (a) Plots of $\gamma$ and $p_{\text{noise}}$ based on the 2nd to 4th order correlators, where $n = 10$, $m = 15$ and the individual sample number is $2 \times 10^4$. (b) The clouds based on the 3rd order correlators with changing single-photon input combinations, where $n = 4$, $m = 16$ and the individual sample number is $10^4$. (c) Plots of mean CS (CV) and $p_{\text{noise}}$ according to (b).

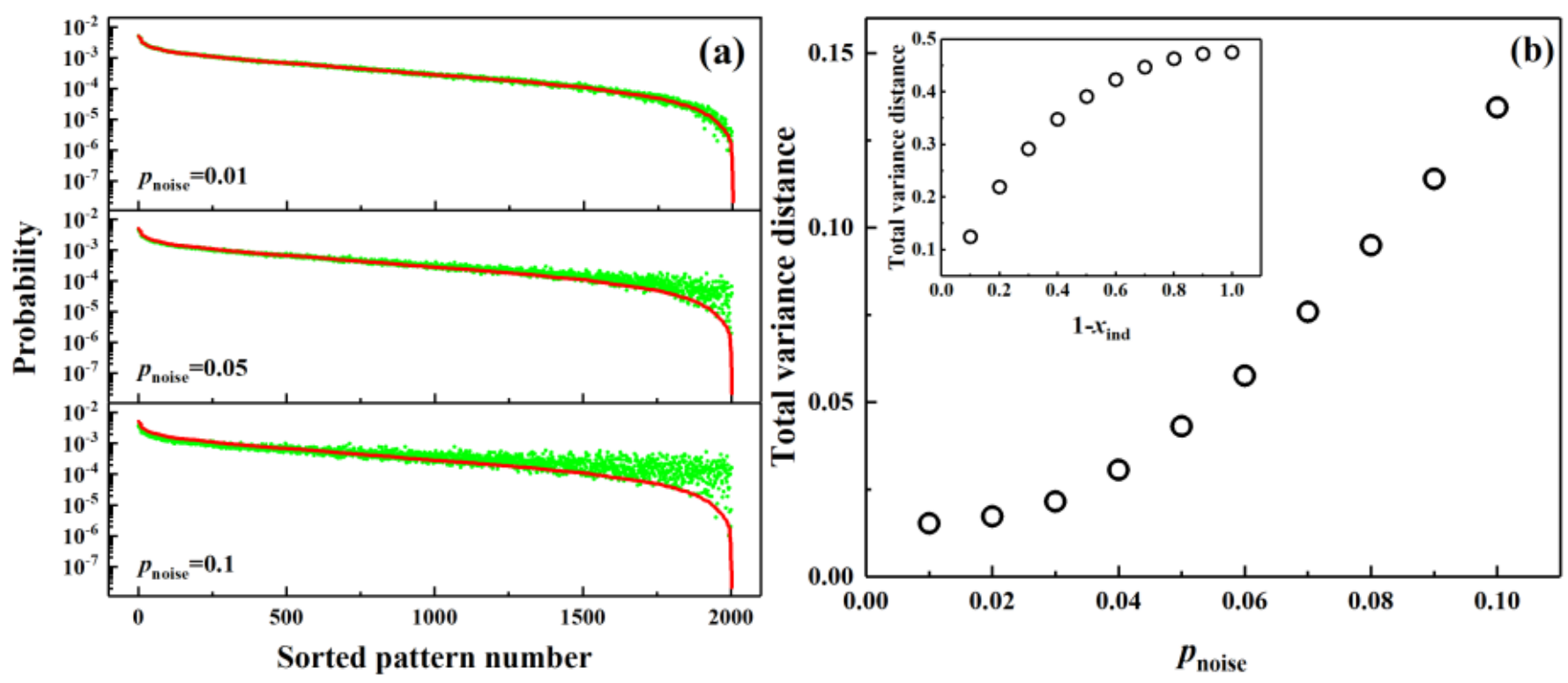

Fig. 8 Output probability deviations with noises. (a) Distributions of sorted output patterns with $p_{\text{noise}}$. The red lines show the background distributions with $n = 5$ and $m = 10$, where the patterns are sorted according to probabilities from high to low. The green dots present distributions of the correspondent patterns changed with $p_{\text{noise}}$, which are obtained approximately by simulating $10^6$ samples for each $p_{\text{noise}}$. (b) Plots of the correspondent total variance distance and $p_{\text{noise}}$. The inset shows plots with changing $x_{\text{ind}}$.

The extended statistical benchmark methods are also employed to evaluate the noises of photon loss compensated by dark counts, with input and output photon numbers kept equal. For simplicity, we keep the condition that $1 - \eta_{\text{t}} = p_{\text{loss}} = p_{\text{dc}} = p_{\text{noise}}$. Note that in general, $p_{\text{dc}}$ is quite low and the condition is not realistic. This extreme condition is considered to test performances of statistical benchmark methods. The noise evaluations are presented in Fig. 7. With enhancing the noise level, $\gamma$ decreases in tendency, whereas some points may be deviated from the tendency especially when the noise level is quite low. As for the cloud method, the changes of mean CS and CV are continuous with the noise level.

It is notable that in Fig. 7, when the noise level is quite low, the evaluations go through poor performances with drastic fluctuations or inapparent changes, which is totally contrast to the photon partial distinguishability case. The influences of $p_{\text{noise}}$ are further investigated by observing the distributions of sorted output patterns. The probability calculations of all the output patterns are relatively difficult because of numerous combinations when $p_{\text{noise}}$ is not zero. However, we can obtain the distributions approximately by simulating massive amounts of samples since the simulations are fast with fixed background probabilities. The distributions in Fig. 8 (a) indicate that only when $p_{\text{noise}}$ is relatively high, probability deviations are obvious, leading to the disappearance of unbalances. Indeed, it can be understood by considering the process that the unbalanced distribution is first partly eliminated by photon loss, and then replaced by a more balanced one due to dark counts.

Deviations can also be evaluated by the total variance distance $D = \frac{1}{2}\sum_T |\text{Pr}_{\text{noise}}(T) -$

$\mathrm{Pr}_{\mathrm{ideal}}(T)|$ [21], where $\mathrm{Pr}_{\mathrm{noise}}(T)$ and $\mathrm{Pr}_{\mathrm{ideal}}(T)$ are probabilities of the output pattern $T$ with and without noises. As shown in Fig. 8 (b), the enhancement of $D$ is first suppressed when $p_{\mathrm{noise}}$ is quite low, in which region the statistical benchmark methods have poor performances. As a contrast, as for the noise of photon partial distinguishability, when $x_{\mathrm{ind}}$ is slightly deviated from 1, the enhancement of $D$ is obvious. This indicates once again that keeping photons indistinguishable is very important in boson sampling.

# 4 Conclusion

In conclusion, noises in boson sampling are quantificationally evaluated with using statistical benchmark methods, which are computational efficient. These methods are found to be more effective when higher order correlators are used, which are more sensitive to multi-photon interferences. The cloud method performs well, working in the regime $m \gg n$ and free from learning interferometer information. It is notable that the noises of photon loss compensated by dark counts can also lead to the disappearance of output distribution unbalances. An effective scheme to fast simulate noisy boson samples are also provided, especially those samples with the essential noise of photon partial distinguishability.

***Acknowledgements.*** This work can be carried out thanks to Project supported by Shanghai Municipal Science and Technology Major Project (Grant No. 2019SHZDZX01).